\documentclass[aip,graphicx,preprint]{revtex4-1}

\usepackage[english]{babel}
\usepackage{graphicx}% Include figure files
\usepackage{dcolumn}% Align table columns on decimal point
\usepackage{bm}
\usepackage{upgreek}

\begin{document}

%\preprint{}


%\preprint{}

\title{High-Responsivity Graphene-Boron Nitride Photodetector and Autocorrelator in a Silicon Photonic Integrated Circuit}% 

\author{Ren-Jye Shiue}

\thanks{These authors contribute equally to this work}

\affiliation{Department of Electrical Engineering and Computer Science, Massachusetts Institute of Technology}

\author{Yuanda Gao}
%\altaffiliation{These authors contribute equally to this work}
\thanks{These authors contribute equally to this work}

\affiliation{Department of Mechanical Engineering, Columbia University}

\author{Yifei Wang}
\thanks{These authors contribute equally to this work}

\affiliation{Department of Electrical Engineering and Computer Science, Massachusetts Institute of Technology}

\author{Cheng Peng}
\affiliation{Department of Electrical Engineering and Computer Science, Massachusetts Institute of Technology}

\author{Alexander D. Robertson}
\affiliation{Department of Mechanical Engineering, Columbia University}

\author{Dmitri K. Efetov}
\affiliation{Department of Electrical Engineering and Computer Science, Massachusetts Institute of Technology}

\author{Solomon Assefa}
\affiliation{IBM T. J. Watson Research Center}

\author{Frank H. L. Koppens}
\affiliation{ICFO - Institut de Ciencies Fotoniques}

\author{James Hone}
\affiliation{Department of Mechanical Engineering, Columbia University}

\author{Dirk Englund}
\email{englund@mit.edu}
\affiliation
{Department of Electrical Engineering and Computer Science, Massachusetts Institute of Technology}




\date{\today}

\begin{abstract}

Graphene and other two-dimensional (2D) materials have emerged as promising materials for broadband and ultrafast photodetection and optical modulation. These optoelectronic capabilities can augment complementary metal-oxide-semiconductor (CMOS) devices for high-speed and low-power optical interconnects. Here, we demonstrate an on-chip ultrafast photodetector based on a two-dimensional heterostructure consisting of high-quality graphene encapsulated in hexagonal boron nitride. Coupled to the optical mode of a silicon waveguide, this 2D heterostructure-based photodetector exhibits a maximum responsivity of 0.36 A/W and high-speed operation with a 3 dB cut-off at 42 GHz. From photocurrent measurements as a function of the top-gate and source-drain voltages, we conclude that the photoresponse is consistent with hot electron mediated effects. At moderate peak powers above 50~mW, we observe a saturating photocurrent consistent with the mechanisms of electron-phonon supercollision cooling. This nonlinear photoresponse enables optical on-chip autocorrelation measurements with picosecond-scale timing resolution and exceptionally low peak powers.

\end{abstract}


\maketitle

The past decades have seen concerted efforts to combine optical and electrical components into integrated optoelectronic circuits for a variety of applications, including optical interconnects~\cite{Hochberg2010,2009.Miller,2014.Book.Springer.Bergman_et_al.photonic_networks_on_chip}, biochemical sensing~\cite{Estevez2012}, and all-optical signal processing\cite{2015.Book.Springer.Eggleton.all-optical-signal}. The leading photodetector architectures, which form an essential technology for such circuits,  are presently based on bulk semiconductor materials, predominantly Ge~\cite{Wang2011p} and InP\cite{Liang2010}. However, the integration of bulk semiconductor-based on-chip detectors has faced important challenges, including in some cases front-end changes in complementary metal-oxide-semiconductor (CMOS) processing, a spectral response limited by the material's bandgap---e.g. up to the L telecommunication band for strained Ge---and an intrinsic speed limited by the material's carrier mobility. In addition, because of the weak nonlinear optical response of these bulk semiconductor materials, high optical powers are required in nonlinear applications including optical pulse characterization, sampling, and signal processing\cite{Slavik2010,Ferdous2011,Weiner2000}. Recently, the family of two-dimensional (2D) materials has emerged as an alternative optoelectronic platform with new functionalities\cite{Koppens2014,Avouris2014a}, including broadband ultrafast photodetection\cite{Xia2014,2013.IEEE.Gan.graphene-photonics,Youngblood2015,2014.ACS.Schall.50Gbps_graphene,Tielrooij2015}, on-chip electro-optic modulation\cite{Liu2011d,2015.NanoLett.Gao.2D_PC_graphene_modulator,2014.ArXiv.Lipson.graphene_modulator,Hu2014}, light emission\cite{Baugher2014,Pospischil2014}, saturable absorption\cite{Sun2010}, and parametric nonlinearities\cite{Gu2012,Hendry2010a,Gullans2013}. Moreover, distinct 2D materials can be assembled nearly defect-free into entirely new types of 2D heterostructures\cite{Wang2013b,Geim2013}, enabling  optoelectronic properties and device concepts that were not feasible using bulk semiconductors.

\begin{figure}

  \includegraphics[width=17.0cm]{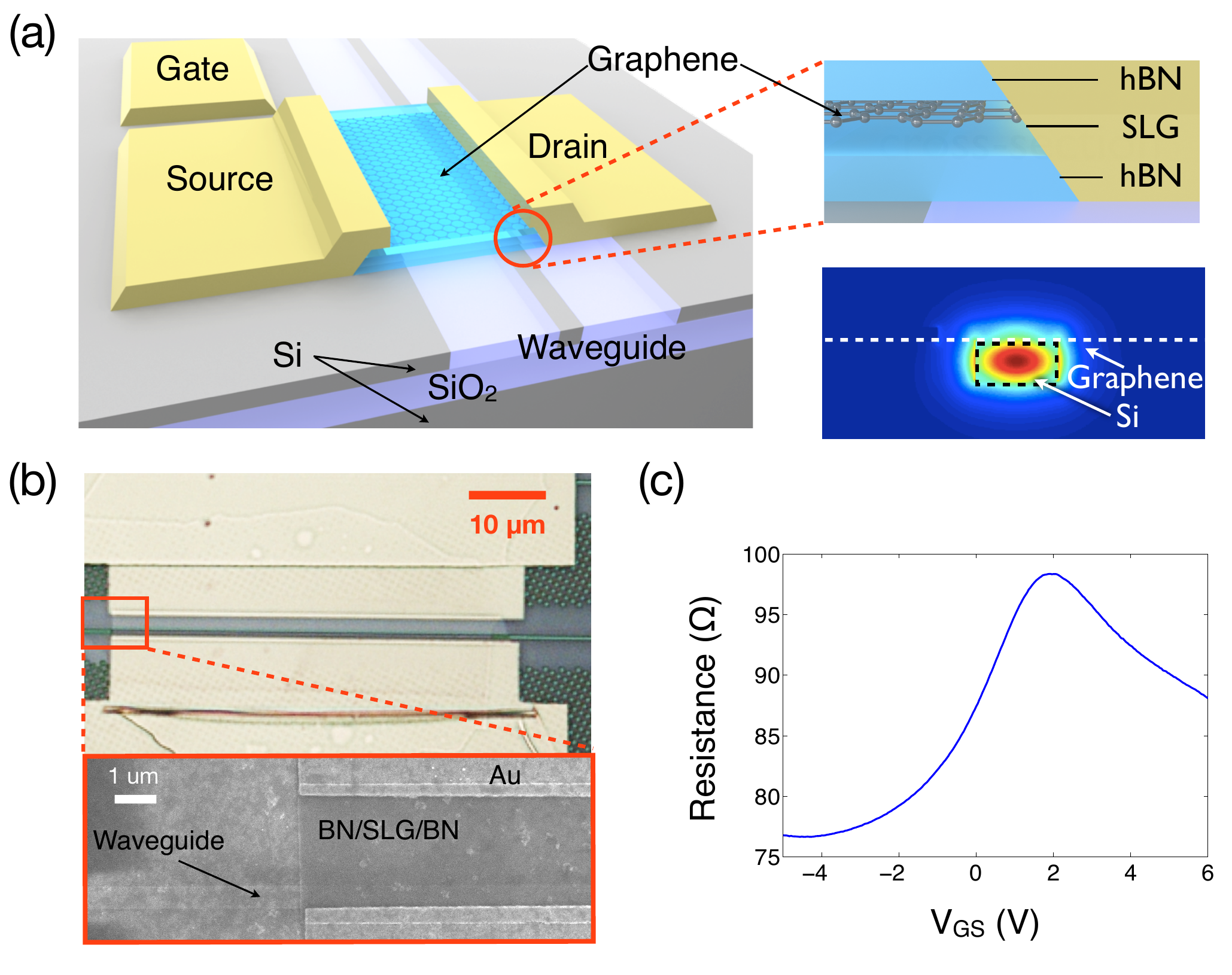}% Here is how to import EPS art
  
  \caption{(a) A schematic of the hBN/SLG/hBN photodetector on a buried silicon waveguide. The lower-right inset shows the electric field energy density of the fundamental TE waveguide mode, obtained by finite-element simulation. (b) Optical microscope image (upper panel) and a scanning electron micrograph (lower panel) of the completed device. (c) Resistance of the graphene device as a function of gate-source voltage V$_{GS}$. A resistance peak at $V_{GS}$ = 2 V indicates the charge neutrality point.}
  
\end{figure}

Here, we introduce a 2D heterostructure consisting of single-layer graphene (SLG) encapsulated by hexagonal boron nitride (hBN) that enables both high-responsivity photodetection and ultrafast pulse metrology in a compact waveguide-integrated design. The mobility of BN-encapsulated graphene can reach up to 80,000 cm$^{2}$/Vs at room temperature, exceeding that of traditional semiconductors such as Ge by 1-3 orders of magnitude. We use one-dimensional contacts to the hBN/SLG/hBN stack combined with a wide channel design to achieve a device resistance as low as $\sim$ 77 $\Omega$. We measure operating speeds exceeding 40 GHz (3dB cut-off) and a record-high detection responsivity exceeding 350 mA/W, approaching responsivities of on-chip Si-Ge photodetectors\cite{Wang2011p}. We fabricated this 2D heterostructure in a back-end-of-the-line (BEOL) step on a silicon-on-oxide (SOI) photonic integrated circuit (PIC), which was fabricated in a CMOS-compatible process. The strong electron-electron interaction and weak electron-phonon coupling of the SLG\cite{Song2011b,Song2012,Graham2012} result in a nonlinear photoresponse above $\sim 50$ mW of peak input power, enabling direct autocorrelation characterization of ultrafast pulses. Unlike traditional free-space autocorrelators, optical nonlinear time-lens techniques\cite{Foster2008}, and frequency-resolved optical gating (FROG)\cite{Tien2009,Walmsley2009}, this on-chip 2D heterostructure-based  autocorrelator does not require parametric nonlinear effects, passive dispersive optics, or separate photodetectors. We measured a minimum timing resolution of 3 ps with a minimum required peak power of 67 mW, comparable to the the autocorrelators based on two photon absorption (TPA) of silicon or III-V compound semiconductors. The ability for high responsivity photodetection  and ultrafast optical autocorrelation in one small-footprint device opens the door to new applications and device concepts in integrated optics.

Fig. 1a illustrates the device architecture. Silicon waveguides were fabricated in an SOI wafer with a 220 nm silicon membrane on a 3-$\upmu$m-thick SiO$_2$ layer. This step used shallow trench isolation (STI) commonly used in CMOS processing. A waveguide width of 520 nm supports a single transverse-electric (TE) mode shown in the simulated mode in Fig. 1a (lower right panel). The chip was planarized by backfilling with a thick SiO$_2$ layer, followed by chemical mechanical polishing to reach the top silicon surface. We then transferred the multilayered hBN/SLG/hBN stack onto the photonic chip using van der Waals (vdW) assembly~\cite{Wang2013b}. The graphene channel spans 40 $\upmu$m of the waveguide. We applied one-dimensional edge contacts to the encapsulated graphene layer using a series of etching and metallization steps\cite{Wang2013b}. A polymer electrolyte (poly(ethylene oxide) and LiClO$_4$) layer was spin-coated on the entire chip to gate the graphene device\cite{Lu2004}. 
 
Fig. 1b shows the completed structure. The drain electrode is positioned only 200~nm from the waveguide to induce a pn junction near the optical mode\cite{Pospischil2013}. Confocal scanning beam excitation measurements confirm a strong photoresponse region near this metal/graphene interface, as demonstrated previously~\cite{Gan2013} and shown for this particular device in Fig. S1 in the supporting information.  Measurements of the device resistance as a function of gate voltage, shown in Fig. 1c, indicate a charge neutrality point (CNP) at a gate-source voltage of $V_{GS}$ = 2 V and a minimum device resistance of 77 $\Omega$. From similar hBN/SLG/hBN stacks assembled on SiO$_2$ substrates, which also included a Hall-bar geometry that was not possible for us to include on the waveguide-integrated device, the mobility of graphene is in the range of 40,000--60,000 cm$^{2}$/V$\cdot$s at carrier densities between 2--4$\times 10^{12}$ cm$^{-2}$\cite{Wang2013b}.

\begin{figure}

  \includegraphics[width=17.0cm]{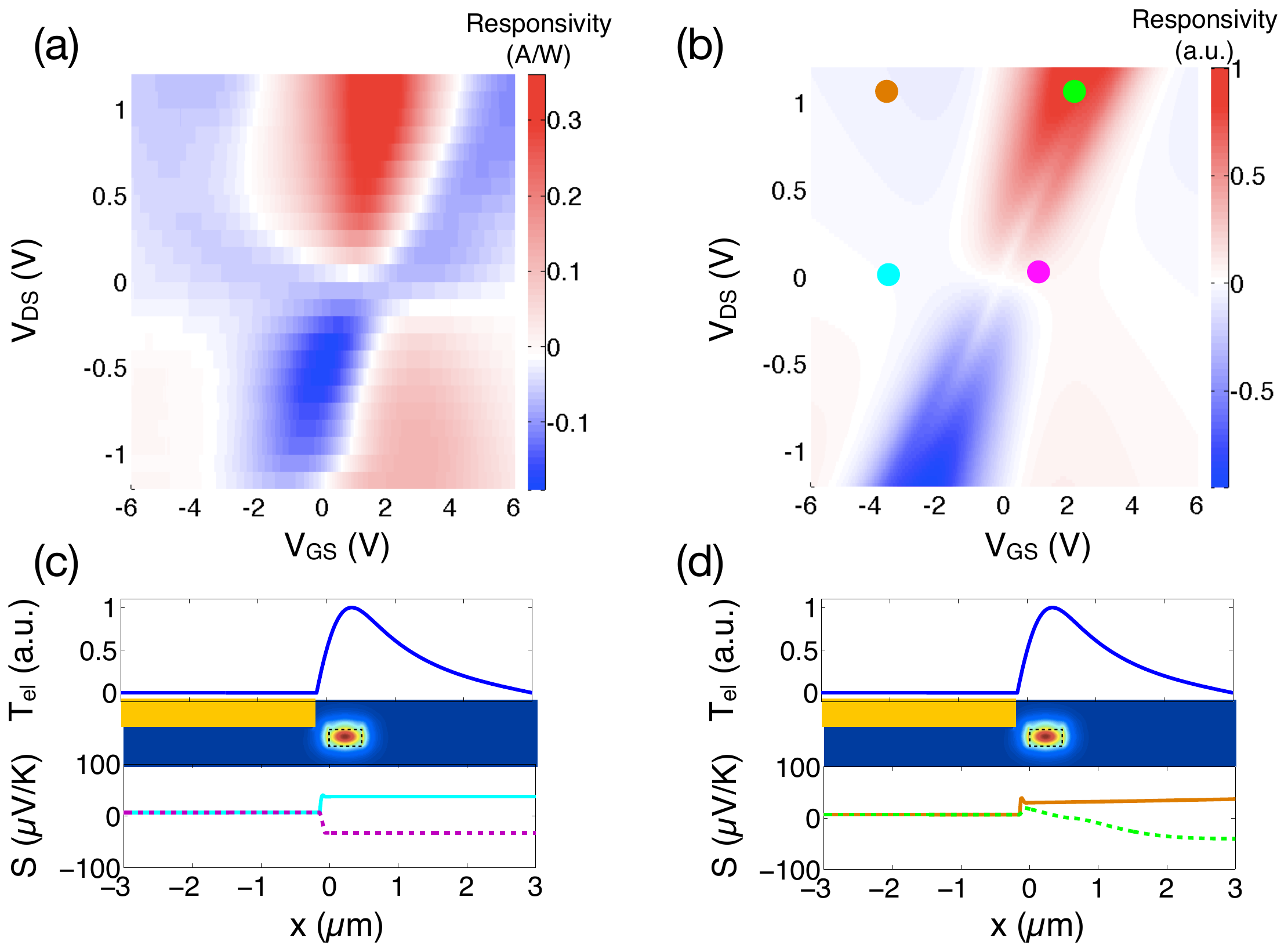}% Here is how to import EPS art
  
  \caption{(a) Measured and (b) calculated responsivity as a function of gate-source $V_{GS}$ and drain-source $V_{DS}$ voltages. (c) and (d) shows the electron temperature profile of the SLG under c.w. optical excitation from the waveguide (top panel). Bottom panel: The Seebeck coefficient profiles of the SLG at different V$_{GS}$ and V$_{DS}$ that are indicated as the colored dots in (b).}

\end{figure}

We characterized the transmission of the completed device in the telecommunications band at 1550 nm, using lensed fibers for edge-coupling to SiN mode converters at the input and output of the $\sim 4$-mm-long Si waveguide, as reported previously~\cite{Gan2013}. The hBN/SLG/hBN detector stack is located at the center of the waveguide, $\sim$ 2 mm from either edge facet. Depositing the stack increased the waveguide transmission loss from 11 dB to 13.2 dB. Attributing this 2.2 dB excess loss to the 40-$\upmu$m-long graphene-waveguide overlap, we calculate an absorption coefficient of 0.055 dB/$\upmu$m. This coefficient is slightly higher than the absorption of 0.043 dB/$\upmu$m that we estimated independently from a finite element simulation of the waveguide evanescent field coupling to the SLG\cite{Gan2013} (note that the hBN is transparent in the telecom band since it has a bandgap of 5.2 eV). We attribute the slightly higher loss (0.48 dB) of the experimental device to additional scattering and reflection losses at the waveguide-detector stack interface. We observed negligible  transmission loss from the electrolyte.

Fig. 2a presents responsivity measurements as a function of $V_{GS}$ and $V_{DS}$ when the detector is illuminated via the waveguide at 25 $\upmu$W continuous-wave (c.w.) laser inside the waveguide. Here the responsivity is defined as the ratio of the short-circuit photocurrent (supporting information) to the optical power in the waveguide, $R = I_{ph}/P_{in}$. Unlike previous waveguide-integrated graphene photodetectors, which included only drain-source voltage control~\cite{Gan2013,Pospischil2013,Wang2013}, the additional top electrolyte-gate allows us to independently tune the graphene Fermi level and electric field across the waveguide mode. Scanning $V_{GS}$ and $V_{DS}$ produces a six-fold pattern in the photocurrent, which qualitatively matches the behavior of the photothermoelectric (PTE) effect\cite{Song2011b,Gabor2011c}. The photocurrent reaches a maximum of 0.36 A/W at V$_{GS}$ = 2 V and V$_{DS}$ = 1.2 V; this represent a more than three-fold improvement over previous waveguide-integrated graphene photodetectors\cite{Gan2013,Pospischil2013,Wang2013}.

To explain the measured six-fold pattern in the responsivity map, we consider the PTE effect for the metal/graphene junction that overlaps with the waveguide. The evanescent field from the waveguide creates photo-excited carriers in the SLG, which rapidly thermalize, increasing the local electron temperature. Fig. 2c (top panel) shows the numerically solved local electron temperature $T_{el}$ across the source and drain contacts of the detector (supporting information). The photo-excited hot-electrons diffuse and create a potential gradient ${\Delta}V=-S(x){\nabla}T_{el}(x)$, where ${\nabla}T_{el}(x)$ is the temperature gradient of the hot electrons and $S(x)$ is the Seebeck coefficient. The total photocurrent collected between the source and drain contacts is given by

\begin{equation}
I_{Ph}=G\int_{0}^{L} S(x)\nabla T(x)dx
\end{equation}

\noindent where $G$ is the conductance of graphene and $L$ the is length of the graphene channel. From the Mott formula\cite{Ashcroft1976,Liu2012,Lemme2011,Song2011b}, $S$ can be expressed as

\begin{equation}
S(\mu)=-\frac{\pi^2k_B^2T}{3e}\frac{1}{\sigma}\frac{d\sigma}{d\mu^\prime}\end{equation}

\noindent where $k_B$ is the Boltzmann constant, $T$ is the lattice temperature, $e$ is the electron charge, $\sigma$ is the conductivity of graphene, and $\mu$ is the chemical potential. Eq. 2 indicates that $S$ depends strongly on the carrier density and the chemical potential of graphene, which can be tuned by $V_{GS}$ and $V_{DS}$ in our device. Fig. 2c and 2d show examples of the spatially resolved S(x) using parameters obtained from the measured conductance and the capacitance of the electrolyte (supporting information). By integrating $S(x)$ and $\nabla T_{el}$ along the source and drain contacts using Eq. 1, we can therefore calculate the photocurrent with respect to $V_{GS}$ and $V_{DS}$, producing the photocurrent mapping in Fig. 2b, which shows a six-fold pattern qualitatively similar to the measured gate- and bias-dependent responsivity. It is also important to consider photovoltaic (PV) effects that generate photocurrent at the graphene $p-n$ junction. However, the PV effect only produces a single sign reversal with respect to V$_{GS}$ and V$_{DS}$ and does not agree to the measured six-fold pattern in the experiment\cite{Song2011b,Lemme2011,Liu2012}. We therefore estimate the PV effect to be small compared with the PTE, which is confirmed by theoretical approximations presented in the supporting information.

\begin{figure}

  \includegraphics[width=17.0cm]{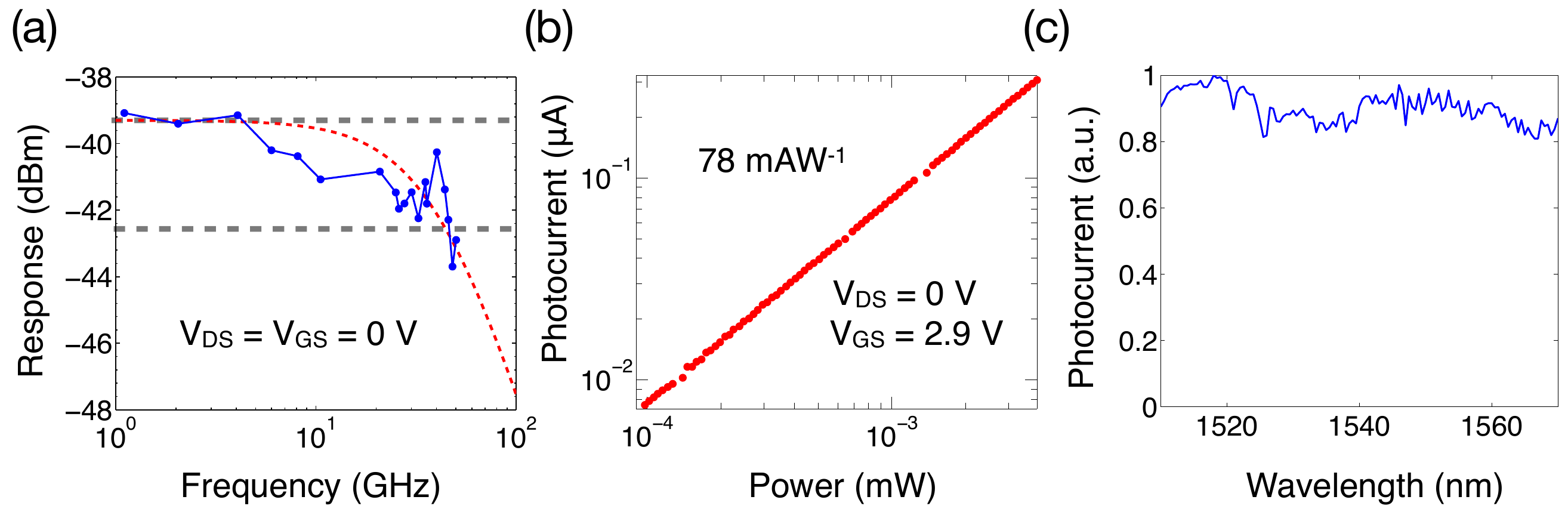}% Here is how to import EPS art
  
  \caption{(a) High-speed response of the graphene photodetector. The red dashed line shows the fitting to the experiment results with a RC low-pass filter model. The extracted 3 dB cut-off frequency is at 42 GHz. (b) Photocurrent as a function of input optical power. (c) Spectral response of the normalized photocurrent for c.w. laser excitation.}
  
\end{figure}

To test the dynamic response of the hBN/SLG/hBN detector, we coupled two narrowband (1 MHz) laser sources with a detuning  frequency $\Delta f$ ranging from 1 to 50 GHz into the waveguide. The interference of these two laser fields produces an intensity in the detector that oscillates at frequency $\Delta f$ . Fig.3a displays the measured power at $\Delta f$, obtained on an electrical spectrum analyzer (maximum frequency 50 GHz) via high-speed RF probes (Cascade FPC-GS-150). This measurement indicates a 3-dB cutoff frequency at 42 GHz, matching the highest reported graphene photodetector speeds~\cite{2014.ACS.Schall.50Gbps_graphene}. We observed this ultrafast photoresponse even under zero drain-source bias, which distinguishes graphene from typical semiconductor high-speed photodiodes\cite{Wang2011p}. The detected photocurrent at $V_{DS}$ = 0 (Fig. 3b) linearly reduces with input power, indicating vanishing dark current. The responsivity at zero bias reaches 78 mA/W. Accounting for the fact that the graphene absorbs only $\sim 1.7-2.2$dB of optical power in the waveguide, we can thus estimate an internal quantum efficiency of 16-19\%. As shown in Fig. 3c, the spectral response of the detector under c.w. laser excitation varies uniformly in the spectral range from 1510 nm to 1570 nm.

\begin{figure}
  \includegraphics[width=17.0cm]{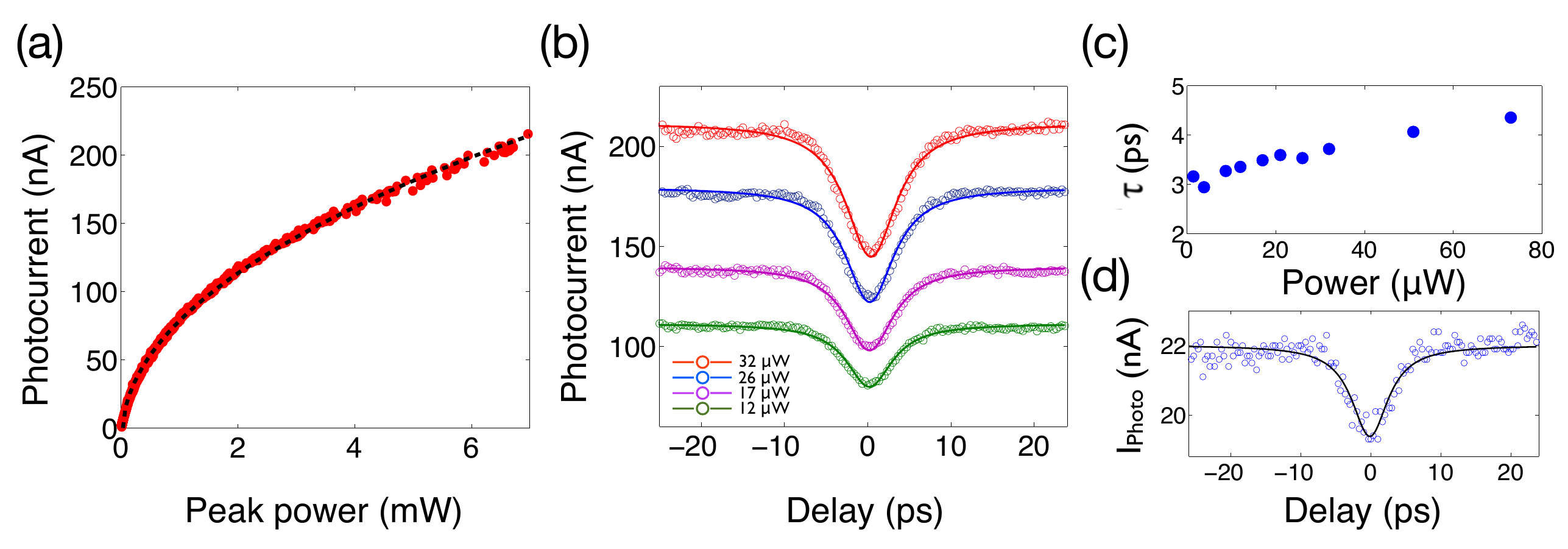}% Here is how to import EPS art
  
  \caption{(a) Photocurrent of the on-chip graphene detector under pulsed (250 fs) excitation. The black curve shows a fitting of the experimental data (red dots) with $I_{Photo} \propto P_{in}^{0.47\pm0.003}$. (b) Autocorrelation traces of the graphene-based autocorrelator with different input average power of the laser pulses. (c) The response time ($\tau$) of the graphene autocorrelator at different input excitation power. (d) An autocorrelation trace measured with small input average power of 1.4 $\upmu$W, corresponding to a peak power of 67 mW.}
  
\end{figure}

We next tested the detector using 250-fs pulses at 1800 nm, produced by an optical parametric oscillator pumped by a Ti:Sapphire oscillator with 78 MHz repetition rate. As shown in Fig. 4a, the photocurrent shows a nonlinear photoresponse that is fit well by $I_{Ph} \propto P_{in}^{0.47\pm0.003}$. This approximately square root dependence in photocurrent with pump power suggests a supercollision-dominated cooling mechanism, for which $I_{Ph} \propto \sqrt{P_{in}}$\cite{Graham2012}. Using this nonlinear photoresponse, we characterized our photodetector's ultimate speed limitations, which are dictated by the internal carrier dynamics.  Fig. 4b plots photocurrent traces as a function of the time delay $\Delta t$ between pairs of 250-fs laser pulses for a range of incident powers. These traces show a clear dip at $\Delta t=0$ with a width that corresponds to the carrier relaxation time, i.e., the time needed for the graphene detector to return to equilibrium. The half-width at half-maximum response times $\tau$ are plotted in Fig. 4c for a sweep of input powers. The minimum response time is approximately $\sim$ 3 ps. The slight increase of the response time at higher input power is unclear to us at this point and requires future studies.

The picosecond-scale nonlinear photocurrent from the photodetector can be used for on-chip autocorrelation measurements for, e.g., ultrafast optical sampling. We tested the autocorrelator with an average power of only 1.4 $\upmu$W (peak power of 67 mW). The autocorrelation trace (Fig. 4d) shows a clear dip with a minimum resolution down to 3ps. The required peak power is less than free space autocorrelators based on parametric frequency conversion \footnote{Reef femtosecond autocorrelator Reef-RTD and Femtochrome Research Inc. FR-103WS}, and is also comparable to TPA-based semiconductor autocorrelators with waveguide or cavity integration\cite{Hayat2011,Duchesne2010,Liang2002}. We noticed that the femtosecond-scale heating of the hot-electrons in graphene in a recent report could in principle improve the timing resolution of the on-chip autocorrelator to sub-50 fs\cite{Tielrooij2015}. It is also important to note that this graphene-based autocorrelator supports a broad spectral range that is identical to the spectral response of the graphene detector, which, in this work, is from 1500 nm to 1800 nm.

In summary, we have demonstrated a compact, high-speed and high-responsivity graphene photodetector integrated into a CMOS-compatible PIC. The maximum responsivity of the detector is 0.36 A/W with a high-speed 42 GHz cut-off frequency. The measured photocurrent of the detector as a function of $V_{GS}$ and $V_{DS}$ can be modeled by the PTE effect at the $p-n$ junction near the metal/graphene interface. Under pulsed excitation, the nonlinear photocurrent response enables the characterization of ultrafast optical pulses and presents a new functionality of such a photodetector for autocorrelation and ultrafast optical sampling measurements. The peak power required for the autocorrelator is as low as 67 mW with a timing resolution of 3 ps. Our work shows the potential of 2D heterostructure-based photodetector architectures integrated in CMOS-compatible fashion with silicon PICs. Moreover, the unique nonlinear photoresponse of graphene enables a versatile platform for ultrafast metrology, which may also find applications for on-chip mode-locked pulse generation and pulse shaping on a single chip. The integrated photodetector and autocorrelator promise a broad range of applications ranging from optical communications to signal processing and ultrafast optics.

%The unique properties of graphene have generated strong interest in developing opto-electronics devices based on the material \cite{Bonaccorso2010,Bao2012}. Examples include graphene-based high speed electro-optical modulators \cite{Liu2011d,Liu2012d}, photodetectors \cite{Mueller2010,Xia2009a}, saturable absorbers \cite{Bao2009,Sun2010}, and nonlinear media for four-wave mixing \cite{Hendry2010,Zhang2011b,Gu2012}. Intrinsic graphene exhibits absorption of 2.3\% \cite{Mak2008b} in the infrared to visible spectra range. While this absorption coefficient is remarkably high for a single atomic layer, for practical applications, a larger absorption coefficient is needed. To increase the light-matter interactions in graphene, approaches to date have included the integration of graphene with optical micro-cavities \cite{Furchia,Majumdar2013,Engel2012,Gan2014,Gan2013a}, plasmonic nanostructures \cite{Echtermeyer2011,Fang2012,Fang2012,Liu2011b,Chen2012l,Fei2012} and silicon photonic waveguides \cite{Liu2011d,Liu2012d,Koester2012,Li2012e}. 

\begin{acknowledgments}

The authors thank Prof. Xuetao Gan for helpful discussion on the experiment scheme and data analysis. Financial support was provided by the Office of Naval Research (Awards N00014-13-1-0662 and N00014-14-1-0349). Device fabrication was partly carried out at the Center for Functional Nanomaterials, Brookhaven National Laboratory, which is supported by the US Department of Energy, Office of Basic Energy Sciences (contract no. DE-AC02- 98CH10886). R.-J.S. was supported in part by the Center for Excitonics, an Energy Frontier Research Center funded by the US Department of Energy, Office of Science, Office of Basic Energy Sciences under award no. DE-SC0001088. YF. W. was supported in part by Tsinghua Xuetang Talents Program.

\end{acknowledgments}


\begin{thebibliography}{47}%
\makeatletter
\providecommand \@ifxundefined [1]{%
 \@ifx{#1\undefined}
}%
\providecommand \@ifnum [1]{%
 \ifnum #1\expandafter \@firstoftwo
 \else \expandafter \@secondoftwo
 \fi
}%
\providecommand \@ifx [1]{%
 \ifx #1\expandafter \@firstoftwo
 \else \expandafter \@secondoftwo
 \fi
}%
\providecommand \natexlab [1]{#1}%
\providecommand \enquote  [1]{``#1''}%
\providecommand \bibnamefont  [1]{#1}%
\providecommand \bibfnamefont [1]{#1}%
\providecommand \citenamefont [1]{#1}%
\providecommand \href@noop [0]{\@secondoftwo}%
\providecommand \href [0]{\begingroup \@sanitize@url \@href}%
\providecommand \@href[1]{\@@startlink{#1}\@@href}%
\providecommand \@@href[1]{\endgroup#1\@@endlink}%
\providecommand \@sanitize@url [0]{\catcode `\\12\catcode `\$12\catcode
  `\&12\catcode `\#12\catcode `\^12\catcode `\_12\catcode `\%12\relax}%
\providecommand \@@startlink[1]{}%
\providecommand \@@endlink[0]{}%
\providecommand \url  [0]{\begingroup\@sanitize@url \@url }%
\providecommand \@url [1]{\endgroup\@href {#1}{\urlprefix }}%
\providecommand \urlprefix  [0]{URL }%
\providecommand \Eprint [0]{\href }%
\providecommand \doibase [0]{http://dx.doi.org/}%
\providecommand \selectlanguage [0]{\@gobble}%
\providecommand \bibinfo  [0]{\@secondoftwo}%
\providecommand \bibfield  [0]{\@secondoftwo}%
\providecommand \translation [1]{[#1]}%
\providecommand \BibitemOpen [0]{}%
\providecommand \bibitemStop [0]{}%
\providecommand \bibitemNoStop [0]{.\EOS\space}%
\providecommand \EOS [0]{\spacefactor3000\relax}%
\providecommand \BibitemShut  [1]{\csname bibitem#1\endcsname}%
\let\auto@bib@innerbib\@empty
%</preamble>
\bibitem [{\citenamefont {Hochberg}\ and\ \citenamefont
  {Baehr-Jones}(2010)}]{Hochberg2010}%
  \BibitemOpen
  \bibfield  {author} {\bibinfo {author} {\bibfnamefont {M.}~\bibnamefont
  {Hochberg}}\ and\ \bibinfo {author} {\bibfnamefont {T.}~\bibnamefont
  {Baehr-Jones}},\ }\bibfield  {title} {\enquote {\bibinfo {title} {{Towards
  fabless silicon photonics}},}\ }\href {\doibase 10.1038/nphoton.2010.172}
  {\bibfield  {journal} {\bibinfo  {journal} {Nature Photonics}\ }\textbf
  {\bibinfo {volume} {4}},\ \bibinfo {pages} {492--494} (\bibinfo {year}
  {2010})}\BibitemShut {NoStop}%
\bibitem [{\citenamefont {Miller}(2009)}]{2009.Miller}%
  \BibitemOpen
  \bibfield  {author} {\bibinfo {author} {\bibfnamefont {D.~A.~B.}\
  \bibnamefont {Miller}},\ }\bibfield  {title} {\enquote {\bibinfo {title}
  {{Device Requirements for Optical Interconnects to Silicon Chips}},}\
  }\href@noop {} {\bibfield  {journal} {\bibinfo  {journal} {Proceedings of the
  IEEE}\ }\textbf {\bibinfo {volume} {97}},\ \bibinfo {pages} {1166--1185}
  (\bibinfo {year} {2009})}\BibitemShut {NoStop}%
\bibitem [{\citenamefont {Bergman}\ \emph {et~al.}(2014)\citenamefont
  {Bergman}, \citenamefont {Carloni}, \citenamefont {Biberman}, \citenamefont
  {Chan},\ and\ \citenamefont
  {Hendry}}]{2014.Book.Springer.Bergman_et_al.photonic_networks_on_chip}%
  \BibitemOpen
  \bibfield  {author} {\bibinfo {author} {\bibfnamefont {K.}~\bibnamefont
  {Bergman}}, \bibinfo {author} {\bibfnamefont {L.~P.}\ \bibnamefont
  {Carloni}}, \bibinfo {author} {\bibfnamefont {A.}~\bibnamefont {Biberman}},
  \bibinfo {author} {\bibfnamefont {J.}~\bibnamefont {Chan}}, \ and\ \bibinfo
  {author} {\bibfnamefont {G.}~\bibnamefont {Hendry}},\ }\href@noop {} {\emph
  {\bibinfo {title} {Photonic Network-on-Chip Design}}}\ (\bibinfo  {publisher}
  {Springer},\ \bibinfo {year} {2014})\BibitemShut {NoStop}%
\bibitem [{\citenamefont {Estevez}, \citenamefont {Alvarez},\ and\
  \citenamefont {Lechuga}(2012)}]{Estevez2012}%
  \BibitemOpen
  \bibfield  {author} {\bibinfo {author} {\bibfnamefont {M.}~\bibnamefont
  {Estevez}}, \bibinfo {author} {\bibfnamefont {M.}~\bibnamefont {Alvarez}}, \
  and\ \bibinfo {author} {\bibfnamefont {L.}~\bibnamefont {Lechuga}},\
  }\bibfield  {title} {\enquote {\bibinfo {title} {{Integrated optical devices
  for lab-on-a-chip biosensing applications}},}\ }\href {\doibase
  10.1002/lpor.201100025} {\bibfield  {journal} {\bibinfo  {journal} {Laser \&
  Photonics Reviews}\ }\textbf {\bibinfo {volume} {6}},\ \bibinfo {pages}
  {463--487} (\bibinfo {year} {2012})}\BibitemShut {NoStop}%
\bibitem [{\citenamefont {Wabnitz}\ and\ \citenamefont
  {Eggleton}(2015)}]{2015.Book.Springer.Eggleton.all-optical-signal}%
  \BibitemOpen
  \bibfield  {author} {\bibinfo {author} {\bibfnamefont {S.}~\bibnamefont
  {Wabnitz}}\ and\ \bibinfo {author} {\bibfnamefont {B.~J.}\ \bibnamefont
  {Eggleton}},\ }\href@noop {} {\emph {\bibinfo {title} {All-Optical Signal
  Processing: Data Communication and Storage Applications}}},\ Vol.\ \bibinfo
  {volume} {194}\ (\bibinfo  {publisher} {Springer},\ \bibinfo {year}
  {2015})\BibitemShut {NoStop}%
\bibitem [{\citenamefont {Wang}\ and\ \citenamefont {Lee}(2011)}]{Wang2011p}%
  \BibitemOpen
  \bibfield  {author} {\bibinfo {author} {\bibfnamefont {J.}~\bibnamefont
  {Wang}}\ and\ \bibinfo {author} {\bibfnamefont {S.}~\bibnamefont {Lee}},\
  }\bibfield  {title} {\enquote {\bibinfo {title} {{Ge-photodetectors for
  Si-based optoelectronic integration.}}}\ }\href {\doibase 10.3390/s110100696}
  {\bibfield  {journal} {\bibinfo  {journal} {Sensors (Basel, Switzerland)}\
  }\textbf {\bibinfo {volume} {11}},\ \bibinfo {pages} {696--718} (\bibinfo
  {year} {2011})}\BibitemShut {NoStop}%
\bibitem [{\citenamefont {Liang}\ \emph {et~al.}(2010)\citenamefont {Liang},
  \citenamefont {Roelkens}, \citenamefont {Baets},\ and\ \citenamefont
  {Bowers}}]{Liang2010}%
  \BibitemOpen
  \bibfield  {author} {\bibinfo {author} {\bibfnamefont {D.}~\bibnamefont
  {Liang}}, \bibinfo {author} {\bibfnamefont {G.}~\bibnamefont {Roelkens}},
  \bibinfo {author} {\bibfnamefont {R.}~\bibnamefont {Baets}}, \ and\ \bibinfo
  {author} {\bibfnamefont {J.~E.}\ \bibnamefont {Bowers}},\ }\bibfield  {title}
  {\enquote {\bibinfo {title} {{Hybrid Integrated Platforms for Silicon
  Photonics}},}\ }\href {\doibase 10.3390/ma3031782} {\bibfield  {journal}
  {\bibinfo  {journal} {Materials}\ }\textbf {\bibinfo {volume} {3}},\ \bibinfo
  {pages} {1782--1802} (\bibinfo {year} {2010})}\BibitemShut {NoStop}%
\bibitem [{\citenamefont {Slav\'{\i}k}\ \emph {et~al.}(2010)\citenamefont
  {Slav\'{\i}k}, \citenamefont {Parmigiani}, \citenamefont {Kakande},
  \citenamefont {Lundstr\"{o}m}, \citenamefont {Sj\"{o}din}, \citenamefont
  {Andrekson}, \citenamefont {Weerasuriya}, \citenamefont {Sygletos},
  \citenamefont {Ellis}, \citenamefont {Gr\"{u}ner-Nielsen}, \citenamefont
  {Jakobsen}, \citenamefont {Herstr\o~m}, \citenamefont {Phelan}, \citenamefont
  {O'Gorman}, \citenamefont {Bogris}, \citenamefont {Syvridis}, \citenamefont
  {Dasgupta}, \citenamefont {Petropoulos},\ and\ \citenamefont
  {Richardson}}]{Slavik2010}%
  \BibitemOpen
  \bibfield  {author} {\bibinfo {author} {\bibfnamefont {R.}~\bibnamefont
  {Slav\'{\i}k}}, \bibinfo {author} {\bibfnamefont {F.}~\bibnamefont
  {Parmigiani}}, \bibinfo {author} {\bibfnamefont {J.}~\bibnamefont {Kakande}},
  \bibinfo {author} {\bibfnamefont {C.}~\bibnamefont {Lundstr\"{o}m}}, \bibinfo
  {author} {\bibfnamefont {M.}~\bibnamefont {Sj\"{o}din}}, \bibinfo {author}
  {\bibfnamefont {P.~A.}\ \bibnamefont {Andrekson}}, \bibinfo {author}
  {\bibfnamefont {R.}~\bibnamefont {Weerasuriya}}, \bibinfo {author}
  {\bibfnamefont {S.}~\bibnamefont {Sygletos}}, \bibinfo {author}
  {\bibfnamefont {A.~D.}\ \bibnamefont {Ellis}}, \bibinfo {author}
  {\bibfnamefont {L.}~\bibnamefont {Gr\"{u}ner-Nielsen}}, \bibinfo {author}
  {\bibfnamefont {D.}~\bibnamefont {Jakobsen}}, \bibinfo {author}
  {\bibfnamefont {S.~r.}\ \bibnamefont {Herstr\o~m}}, \bibinfo {author}
  {\bibfnamefont {R.}~\bibnamefont {Phelan}}, \bibinfo {author} {\bibfnamefont
  {J.}~\bibnamefont {O'Gorman}}, \bibinfo {author} {\bibfnamefont
  {A.}~\bibnamefont {Bogris}}, \bibinfo {author} {\bibfnamefont
  {D.}~\bibnamefont {Syvridis}}, \bibinfo {author} {\bibfnamefont
  {S.}~\bibnamefont {Dasgupta}}, \bibinfo {author} {\bibfnamefont
  {P.}~\bibnamefont {Petropoulos}}, \ and\ \bibinfo {author} {\bibfnamefont
  {D.~J.}\ \bibnamefont {Richardson}},\ }\bibfield  {title} {\enquote {\bibinfo
  {title} {{All-optical phase and amplitude regenerator for next-generation
  telecommunications systems}},}\ }\href {\doibase 10.1038/nphoton.2010.203}
  {\bibfield  {journal} {\bibinfo  {journal} {Nature Photonics}\ }\textbf
  {\bibinfo {volume} {4}},\ \bibinfo {pages} {690--695} (\bibinfo {year}
  {2010})}\BibitemShut {NoStop}%
\bibitem [{\citenamefont {Ferdous}\ \emph {et~al.}(2011)\citenamefont
  {Ferdous}, \citenamefont {Miao}, \citenamefont {Leaird},\ and\ \citenamefont
  {Srinivasan}}]{Ferdous2011}%
  \BibitemOpen
  \bibfield  {author} {\bibinfo {author} {\bibfnamefont {F.}~\bibnamefont
  {Ferdous}}, \bibinfo {author} {\bibfnamefont {H.}~\bibnamefont {Miao}},
  \bibinfo {author} {\bibfnamefont {D.}~\bibnamefont {Leaird}}, \ and\ \bibinfo
  {author} {\bibfnamefont {K.}~\bibnamefont {Srinivasan}},\ }\bibfield  {title}
  {\enquote {\bibinfo {title} {{Spectral line-by-line pulse shaping of on-chip
  microresonator frequency combs}},}\ }\href {\doibase
  10.1038/NPHOTON.2011.255} {\bibfield  {journal} {\bibinfo  {journal} {Nature
  \ldots}\ }\textbf {\bibinfo {volume} {5}},\ \bibinfo {pages} {770--776}
  (\bibinfo {year} {2011})}\BibitemShut {NoStop}%
\bibitem [{\citenamefont {Weiner}(2000)}]{Weiner2000}%
  \BibitemOpen
  \bibfield  {author} {\bibinfo {author} {\bibfnamefont {A.~M.}\ \bibnamefont
  {Weiner}},\ }\bibfield  {title} {\enquote {\bibinfo {title} {{Femtosecond
  pulse shaping using spatial light modulators}},}\ }\href {\doibase
  10.1063/1.1150614} {\bibfield  {journal} {\bibinfo  {journal} {Review of
  Scientific Instruments}\ }\textbf {\bibinfo {volume} {71}},\ \bibinfo {pages}
  {1929} (\bibinfo {year} {2000})}\BibitemShut {NoStop}%
\bibitem [{\citenamefont {Koppens}\ \emph {et~al.}(2014)\citenamefont
  {Koppens}, \citenamefont {Mueller}, \citenamefont {Avouris}, \citenamefont
  {Ferrari}, \citenamefont {Vitiello},\ and\ \citenamefont
  {Polini}}]{Koppens2014}%
  \BibitemOpen
  \bibfield  {author} {\bibinfo {author} {\bibfnamefont {F.~H.~L.}\
  \bibnamefont {Koppens}}, \bibinfo {author} {\bibfnamefont {T.}~\bibnamefont
  {Mueller}}, \bibinfo {author} {\bibfnamefont {P.}~\bibnamefont {Avouris}},
  \bibinfo {author} {\bibfnamefont {a.~C.}\ \bibnamefont {Ferrari}}, \bibinfo
  {author} {\bibfnamefont {M.~S.}\ \bibnamefont {Vitiello}}, \ and\ \bibinfo
  {author} {\bibfnamefont {M.}~\bibnamefont {Polini}},\ }\bibfield  {title}
  {\enquote {\bibinfo {title} {{Photodetectors based on graphene, other
  two-dimensional materials and hybrid systems}},}\ }\href {\doibase
  10.1038/nnano.2014.215} {\bibfield  {journal} {\bibinfo  {journal} {Nature
  Nanotechnology}\ }\textbf {\bibinfo {volume} {9}},\ \bibinfo {pages}
  {780--793} (\bibinfo {year} {2014})}\BibitemShut {NoStop}%
\bibitem [{\citenamefont {Avouris}\ and\ \citenamefont
  {Freitag}(2014)}]{Avouris2014a}%
  \BibitemOpen
  \bibfield  {author} {\bibinfo {author} {\bibfnamefont {P.}~\bibnamefont
  {Avouris}}\ and\ \bibinfo {author} {\bibfnamefont {M.}~\bibnamefont
  {Freitag}},\ }\bibfield  {title} {{\selectlanguage {english}\enquote
  {\bibinfo {title} {{Graphene Photonics, Plasmonics, and Optoelectronics}},}\
  }}\href {\doibase 10.1109/JSTQE.2013.2272315} {\bibfield  {journal} {\bibinfo
   {journal} {IEEE Journal of Selected Topics in Quantum Electronics}\ }\textbf
  {\bibinfo {volume} {20}},\ \bibinfo {pages} {6000112--6000112} (\bibinfo
  {year} {2014})}\BibitemShut {NoStop}%
\bibitem [{\citenamefont {Xia}\ \emph {et~al.}(2014)\citenamefont {Xia},
  \citenamefont {Wang}, \citenamefont {Xiao}, \citenamefont {Dubey},\ and\
  \citenamefont {Ramasubramaniam}}]{Xia2014}%
  \BibitemOpen
  \bibfield  {author} {\bibinfo {author} {\bibfnamefont {F.}~\bibnamefont
  {Xia}}, \bibinfo {author} {\bibfnamefont {H.}~\bibnamefont {Wang}}, \bibinfo
  {author} {\bibfnamefont {D.}~\bibnamefont {Xiao}}, \bibinfo {author}
  {\bibfnamefont {M.}~\bibnamefont {Dubey}}, \ and\ \bibinfo {author}
  {\bibfnamefont {A.}~\bibnamefont {Ramasubramaniam}},\ }\bibfield  {title}
  {\enquote {\bibinfo {title} {{Two-dimensional material nanophotonics}},}\
  }\href {\doibase 10.1038/nphoton.2014.271} {\bibfield  {journal} {\bibinfo
  {journal} {Nature Photonics}\ }\textbf {\bibinfo {volume} {8}},\ \bibinfo
  {pages} {899--907} (\bibinfo {year} {2014})}\BibitemShut {NoStop}%
\bibitem [{\citenamefont {Gan}\ \emph {et~al.}(2013{\natexlab{a}})\citenamefont
  {Gan}, \citenamefont {Shiue}, \citenamefont {Gao}, \citenamefont {Assefa},
  \citenamefont {Hone},\ and\ \citenamefont
  {Englund}}]{2013.IEEE.Gan.graphene-photonics}%
  \BibitemOpen
  \bibfield  {author} {\bibinfo {author} {\bibfnamefont {X.}~\bibnamefont
  {Gan}}, \bibinfo {author} {\bibfnamefont {R.-J.}\ \bibnamefont {Shiue}},
  \bibinfo {author} {\bibfnamefont {Y.}~\bibnamefont {Gao}}, \bibinfo {author}
  {\bibfnamefont {S.}~\bibnamefont {Assefa}}, \bibinfo {author} {\bibfnamefont
  {J.}~\bibnamefont {Hone}}, \ and\ \bibinfo {author} {\bibfnamefont
  {D.}~\bibnamefont {Englund}},\ }\bibfield  {title} {\enquote {\bibinfo
  {title} {Controlled light-matter interaction in graphene electrooptic devices
  using nanophotonic cavities and waveguides},}\ }\href {\doibase
  10.1109/JSTQE.2013.2273412} {\bibfield  {journal} {\bibinfo  {journal}
  {Selected Topics in Quantum Electronics, IEEE Journal of}\ }\textbf {\bibinfo
  {volume} {20}},\ \bibinfo {pages} {6000311--6000311} (\bibinfo {year}
  {2013}{\natexlab{a}})}\BibitemShut {NoStop}%
\bibitem [{\citenamefont {Youngblood}\ \emph {et~al.}(2015)\citenamefont
  {Youngblood}, \citenamefont {Chen}, \citenamefont {Koester},\ and\
  \citenamefont {Li}}]{Youngblood2015}%
  \BibitemOpen
  \bibfield  {author} {\bibinfo {author} {\bibfnamefont {N.}~\bibnamefont
  {Youngblood}}, \bibinfo {author} {\bibfnamefont {C.}~\bibnamefont {Chen}},
  \bibinfo {author} {\bibfnamefont {S.~J.}\ \bibnamefont {Koester}}, \ and\
  \bibinfo {author} {\bibfnamefont {M.}~\bibnamefont {Li}},\ }\bibfield
  {title} {\enquote {\bibinfo {title} {{Waveguide-integrated black phosphorus
  photodetector with high responsivity and low dark current}},}\ }\href
  {\doibase 10.1038/nphoton.2015.23} {\bibfield  {journal} {\bibinfo  {journal}
  {Nature Photonics}\ }\textbf {\bibinfo {volume} {9}} (\bibinfo {year}
  {2015}),\ 10.1038/nphoton.2015.23}\BibitemShut {NoStop}%
\bibitem [{\citenamefont {Schall}\ \emph {et~al.}(2014)\citenamefont {Schall},
  \citenamefont {Neumaier}, \citenamefont {Mohsin}, \citenamefont {Chmielak},
  \citenamefont {Bolten}, \citenamefont {Porschatis}, \citenamefont {Prinzen},
  \citenamefont {Matheisen}, \citenamefont {Kuebart}, \citenamefont
  {Junginger}, \citenamefont {Templ}, \citenamefont {Giesecke},\ and\
  \citenamefont {Kurz}}]{2014.ACS.Schall.50Gbps_graphene}%
  \BibitemOpen
  \bibfield  {author} {\bibinfo {author} {\bibfnamefont {D.}~\bibnamefont
  {Schall}}, \bibinfo {author} {\bibfnamefont {D.}~\bibnamefont {Neumaier}},
  \bibinfo {author} {\bibfnamefont {M.}~\bibnamefont {Mohsin}}, \bibinfo
  {author} {\bibfnamefont {B.}~\bibnamefont {Chmielak}}, \bibinfo {author}
  {\bibfnamefont {J.}~\bibnamefont {Bolten}}, \bibinfo {author} {\bibfnamefont
  {C.}~\bibnamefont {Porschatis}}, \bibinfo {author} {\bibfnamefont
  {A.}~\bibnamefont {Prinzen}}, \bibinfo {author} {\bibfnamefont
  {C.}~\bibnamefont {Matheisen}}, \bibinfo {author} {\bibfnamefont
  {W.}~\bibnamefont {Kuebart}}, \bibinfo {author} {\bibfnamefont
  {B.}~\bibnamefont {Junginger}}, \bibinfo {author} {\bibfnamefont
  {W.}~\bibnamefont {Templ}}, \bibinfo {author} {\bibfnamefont {A.~L.}\
  \bibnamefont {Giesecke}}, \ and\ \bibinfo {author} {\bibfnamefont
  {H.}~\bibnamefont {Kurz}},\ }\bibfield  {title} {\enquote {\bibinfo {title}
  {50 gbit/s photodetectors based on wafer-scale graphene for integrated
  silicon photonic communication systems},}\ }\href {\doibase
  10.1021/ph5001605} {\bibfield  {journal} {\bibinfo  {journal} {ACS
  Photonics}\ }\textbf {\bibinfo {volume} {1}},\ \bibinfo {pages} {781--784}
  (\bibinfo {year} {2014})},\ \Eprint
  {http://arxiv.org/abs/http://dx.doi.org/10.1021/ph5001605}
  {http://dx.doi.org/10.1021/ph5001605} \BibitemShut {NoStop}%
\bibitem [{\citenamefont {Tielrooij}\ \emph {et~al.}(2015)\citenamefont
  {Tielrooij}, \citenamefont {Piatkowski}, \citenamefont {Massicotte},
  \citenamefont {Woessner}, \citenamefont {Ma}, \citenamefont {Lee},
  \citenamefont {Myhro}, \citenamefont {Lau}, \citenamefont {Jarillo-Herrero},
  \citenamefont {van Hulst},\ and\ \citenamefont {Koppens}}]{Tielrooij2015}%
  \BibitemOpen
  \bibfield  {author} {\bibinfo {author} {\bibfnamefont {K.~J.}\ \bibnamefont
  {Tielrooij}}, \bibinfo {author} {\bibfnamefont {L.}~\bibnamefont
  {Piatkowski}}, \bibinfo {author} {\bibfnamefont {M.}~\bibnamefont
  {Massicotte}}, \bibinfo {author} {\bibfnamefont {a.}~\bibnamefont
  {Woessner}}, \bibinfo {author} {\bibfnamefont {Q.}~\bibnamefont {Ma}},
  \bibinfo {author} {\bibfnamefont {Y.}~\bibnamefont {Lee}}, \bibinfo {author}
  {\bibfnamefont {K.~S.}\ \bibnamefont {Myhro}}, \bibinfo {author}
  {\bibfnamefont {C.~N.}\ \bibnamefont {Lau}}, \bibinfo {author} {\bibfnamefont
  {P.}~\bibnamefont {Jarillo-Herrero}}, \bibinfo {author} {\bibfnamefont
  {N.~F.}\ \bibnamefont {van Hulst}}, \ and\ \bibinfo {author} {\bibfnamefont
  {F.~H.~L.}\ \bibnamefont {Koppens}},\ }\bibfield  {title} {\enquote {\bibinfo
  {title} {{Generation of photovoltage in graphene on a femtosecond timescale
  through efficient carrier heating}},}\ }\href {\doibase
  10.1038/nnano.2015.54} {\bibfield  {journal} {\bibinfo  {journal} {Nature
  Nanotechnology}\ }\textbf {\bibinfo {volume} {10}},\ \bibinfo {pages}
  {437--443} (\bibinfo {year} {2015})}\BibitemShut {NoStop}%
\bibitem [{\citenamefont {Liu}\ \emph {et~al.}(2011)\citenamefont {Liu},
  \citenamefont {Yin}, \citenamefont {Ulin-Avila}, \citenamefont {Geng},
  \citenamefont {Zentgraf}, \citenamefont {Ju}, \citenamefont {Wang},\ and\
  \citenamefont {Zhang}}]{Liu2011d}%
  \BibitemOpen
  \bibfield  {author} {\bibinfo {author} {\bibfnamefont {M.}~\bibnamefont
  {Liu}}, \bibinfo {author} {\bibfnamefont {X.}~\bibnamefont {Yin}}, \bibinfo
  {author} {\bibfnamefont {E.}~\bibnamefont {Ulin-Avila}}, \bibinfo {author}
  {\bibfnamefont {B.}~\bibnamefont {Geng}}, \bibinfo {author} {\bibfnamefont
  {T.}~\bibnamefont {Zentgraf}}, \bibinfo {author} {\bibfnamefont
  {L.}~\bibnamefont {Ju}}, \bibinfo {author} {\bibfnamefont {F.}~\bibnamefont
  {Wang}}, \ and\ \bibinfo {author} {\bibfnamefont {X.}~\bibnamefont {Zhang}},\
  }\bibfield  {title} {\enquote {\bibinfo {title} {{A graphene-based broadband
  optical modulator.}}}\ }\href {\doibase 10.1038/nature10067} {\bibfield
  {journal} {\bibinfo  {journal} {Nature}\ }\textbf {\bibinfo {volume} {474}},\
  \bibinfo {pages} {64--67} (\bibinfo {year} {2011})}\BibitemShut {NoStop}%
\bibitem [{\citenamefont {Gao}\ \emph {et~al.}(2015)\citenamefont {Gao},
  \citenamefont {Shiue}, \citenamefont {Gan}, \citenamefont {Li}, \citenamefont
  {Peng}, \citenamefont {Meric}, \citenamefont {Wang}, \citenamefont {Szep},
  \citenamefont {Walker}, \citenamefont {Hone},\ and\ \citenamefont
  {Englund}}]{2015.NanoLett.Gao.2D_PC_graphene_modulator}%
  \BibitemOpen
  \bibfield  {author} {\bibinfo {author} {\bibfnamefont {Y.}~\bibnamefont
  {Gao}}, \bibinfo {author} {\bibfnamefont {R.-J.}\ \bibnamefont {Shiue}},
  \bibinfo {author} {\bibfnamefont {X.}~\bibnamefont {Gan}}, \bibinfo {author}
  {\bibfnamefont {L.}~\bibnamefont {Li}}, \bibinfo {author} {\bibfnamefont
  {C.}~\bibnamefont {Peng}}, \bibinfo {author} {\bibfnamefont {I.}~\bibnamefont
  {Meric}}, \bibinfo {author} {\bibfnamefont {L.}~\bibnamefont {Wang}},
  \bibinfo {author} {\bibfnamefont {A.}~\bibnamefont {Szep}}, \bibinfo {author}
  {\bibfnamefont {D.}~\bibnamefont {Walker}}, \bibinfo {author} {\bibfnamefont
  {J.}~\bibnamefont {Hone}}, \ and\ \bibinfo {author} {\bibfnamefont
  {D.}~\bibnamefont {Englund}},\ }\bibfield  {title} {\enquote {\bibinfo
  {title} {High-speed electro-optic modulator integrated with graphene-boron
  nitride heterostructure and photonic crystal nanocavity},}\ }\href {\doibase
  10.1021/nl504860z} {\bibfield  {journal} {\bibinfo  {journal} {Nano Letters}\
  }\textbf {\bibinfo {volume} {15}},\ \bibinfo {pages} {2001--2005} (\bibinfo
  {year} {2015})},\ \bibinfo {note} {pMID: 25700231},\ \Eprint
  {http://arxiv.org/abs/http://dx.doi.org/10.1021/nl504860z}
  {http://dx.doi.org/10.1021/nl504860z} \BibitemShut {NoStop}%
\bibitem [{\citenamefont {Phare}\ \emph {et~al.}(2014)\citenamefont {Phare},
  \citenamefont {Lee}, \citenamefont {Cardenas},\ and\ \citenamefont
  {Lipson}}]{2014.ArXiv.Lipson.graphene_modulator}%
  \BibitemOpen
  \bibfield  {author} {\bibinfo {author} {\bibfnamefont {C.~T.}\ \bibnamefont
  {Phare}}, \bibinfo {author} {\bibfnamefont {Y.-H.~D.}\ \bibnamefont {Lee}},
  \bibinfo {author} {\bibfnamefont {J.}~\bibnamefont {Cardenas}}, \ and\
  \bibinfo {author} {\bibfnamefont {M.}~\bibnamefont {Lipson}},\ }\bibfield
  {title} {\enquote {\bibinfo {title} {30 ghz zeno-based graphene electro-optic
  modulator},}\ }\href@noop {} {\bibfield  {journal} {\bibinfo  {journal}
  {arXiv preprint arXiv:1411.2053}\ } (\bibinfo {year} {2014})}\BibitemShut
  {NoStop}%
\bibitem [{\citenamefont {Hu}\ and\ \citenamefont {Pantouvaki}(2014)}]{Hu2014}%
  \BibitemOpen
  \bibfield  {author} {\bibinfo {author} {\bibfnamefont {Y.}~\bibnamefont
  {Hu}}\ and\ \bibinfo {author} {\bibfnamefont {M.}~\bibnamefont
  {Pantouvaki}},\ }\bibfield  {title} {\enquote {\bibinfo {title} {{Broadband
  10Gb/s Graphene Electro-Absorption Modulator on Silicon for Chip-Level
  Optical Interconnects}},}\ }\href
  {http://photonics.intec.ugent.be/download/pub\_3503.pdf} {\bibfield
  {journal} {\bibinfo  {journal} {\ldots (IEDM), 2014 IEEE \ldots}\ ,\ \bibinfo
  {pages} {128--131}} (\bibinfo {year} {2014})}\BibitemShut {NoStop}%
\bibitem [{\citenamefont {Baugher}\ \emph {et~al.}(2014)\citenamefont
  {Baugher}, \citenamefont {Churchill}, \citenamefont {Yang},\ and\
  \citenamefont {Jarillo-Herrero}}]{Baugher2014}%
  \BibitemOpen
  \bibfield  {author} {\bibinfo {author} {\bibfnamefont {B.~W.~H.}\
  \bibnamefont {Baugher}}, \bibinfo {author} {\bibfnamefont {H.~O.~H.}\
  \bibnamefont {Churchill}}, \bibinfo {author} {\bibfnamefont {Y.}~\bibnamefont
  {Yang}}, \ and\ \bibinfo {author} {\bibfnamefont {P.}~\bibnamefont
  {Jarillo-Herrero}},\ }\bibfield  {title} {\enquote {\bibinfo {title}
  {{Optoelectronic devices based on electrically tunable p-n diodes in a
  monolayer dichalcogenide.}}}\ }\href {\doibase 10.1038/nnano.2014.25}
  {\bibfield  {journal} {\bibinfo  {journal} {Nature nanotechnology}\ }\textbf
  {\bibinfo {volume} {9}},\ \bibinfo {pages} {262--7} (\bibinfo {year}
  {2014})}\BibitemShut {NoStop}%
\bibitem [{\citenamefont {Pospischil}, \citenamefont {Furchi},\ and\
  \citenamefont {Mueller}(2014)}]{Pospischil2014}%
  \BibitemOpen
  \bibfield  {author} {\bibinfo {author} {\bibfnamefont {A.}~\bibnamefont
  {Pospischil}}, \bibinfo {author} {\bibfnamefont {M.~M.}\ \bibnamefont
  {Furchi}}, \ and\ \bibinfo {author} {\bibfnamefont {T.}~\bibnamefont
  {Mueller}},\ }\bibfield  {title} {\enquote {\bibinfo {title} {{Solar-energy
  conversion and light emission in an atomic monolayer p-n diode.}}}\ }\href
  {\doibase 10.1038/nnano.2014.14} {\bibfield  {journal} {\bibinfo  {journal}
  {Nature nanotechnology}\ }\textbf {\bibinfo {volume} {9}},\ \bibinfo {pages}
  {257--61} (\bibinfo {year} {2014})}\BibitemShut {NoStop}%
\bibitem [{\citenamefont {Sun}\ \emph {et~al.}(2010)\citenamefont {Sun},
  \citenamefont {Hasan}, \citenamefont {Torrisi}, \citenamefont {Popa},
  \citenamefont {Privitera}, \citenamefont {Wang}, \citenamefont {Bonaccorso},
  \citenamefont {Basko},\ and\ \citenamefont {Ferrari}}]{Sun2010}%
  \BibitemOpen
  \bibfield  {author} {\bibinfo {author} {\bibfnamefont {Z.}~\bibnamefont
  {Sun}}, \bibinfo {author} {\bibfnamefont {T.}~\bibnamefont {Hasan}}, \bibinfo
  {author} {\bibfnamefont {F.}~\bibnamefont {Torrisi}}, \bibinfo {author}
  {\bibfnamefont {D.}~\bibnamefont {Popa}}, \bibinfo {author} {\bibfnamefont
  {G.}~\bibnamefont {Privitera}}, \bibinfo {author} {\bibfnamefont
  {F.}~\bibnamefont {Wang}}, \bibinfo {author} {\bibfnamefont {F.}~\bibnamefont
  {Bonaccorso}}, \bibinfo {author} {\bibfnamefont {D.~M.}\ \bibnamefont
  {Basko}}, \ and\ \bibinfo {author} {\bibfnamefont {A.~C.}\ \bibnamefont
  {Ferrari}},\ }\bibfield  {title} {\enquote {\bibinfo {title} {{Graphene
  Mode-Locked Ultrafast Laser}},}\ }\href {\doibase 10.1021/nn901703e}
  {\bibfield  {journal} {\bibinfo  {journal} {ACS Nano}\ }\textbf {\bibinfo
  {volume} {4}},\ \bibinfo {pages} {803--810} (\bibinfo {year}
  {2010})}\BibitemShut {NoStop}%
\bibitem [{\citenamefont {Gu}\ \emph {et~al.}(2012)\citenamefont {Gu},
  \citenamefont {Petrone}, \citenamefont {Mcmillan}, \citenamefont {Zande},
  \citenamefont {Yu}, \citenamefont {Lo}, \citenamefont {Kwong}, \citenamefont
  {Hone},\ and\ \citenamefont {Wong}}]{Gu2012}%
  \BibitemOpen
  \bibfield  {author} {\bibinfo {author} {\bibfnamefont {T.}~\bibnamefont
  {Gu}}, \bibinfo {author} {\bibfnamefont {N.}~\bibnamefont {Petrone}},
  \bibinfo {author} {\bibfnamefont {J.~F.}\ \bibnamefont {Mcmillan}}, \bibinfo
  {author} {\bibfnamefont {A.~V.~D.}\ \bibnamefont {Zande}}, \bibinfo {author}
  {\bibfnamefont {M.}~\bibnamefont {Yu}}, \bibinfo {author} {\bibfnamefont
  {G.~Q.}\ \bibnamefont {Lo}}, \bibinfo {author} {\bibfnamefont {D.~L.}\
  \bibnamefont {Kwong}}, \bibinfo {author} {\bibfnamefont {J.}~\bibnamefont
  {Hone}}, \ and\ \bibinfo {author} {\bibfnamefont {C.~W.}\ \bibnamefont
  {Wong}},\ }\bibfield  {title} {\enquote {\bibinfo {title} {{Regenerative
  oscillation and four-wave mixing in graphene optoelectronics}},}\ }\href
  {\doibase 10.1038/NPHOTON.2012.147} {\bibfield  {journal} {\bibinfo
  {journal} {Nature photonics}\ }\textbf {\bibinfo {volume} {43}},\ \bibinfo
  {pages} {1--6} (\bibinfo {year} {2012})}\BibitemShut {NoStop}%
\bibitem [{\citenamefont {Hendry}\ \emph {et~al.}(2010)\citenamefont {Hendry},
  \citenamefont {Hale}, \citenamefont {Moger}, \citenamefont {Savchenko},\ and\
  \citenamefont {Mikhailov}}]{Hendry2010a}%
  \BibitemOpen
  \bibfield  {author} {\bibinfo {author} {\bibfnamefont {E.}~\bibnamefont
  {Hendry}}, \bibinfo {author} {\bibfnamefont {P.~J.}\ \bibnamefont {Hale}},
  \bibinfo {author} {\bibfnamefont {J.}~\bibnamefont {Moger}}, \bibinfo
  {author} {\bibfnamefont {A.~K.}\ \bibnamefont {Savchenko}}, \ and\ \bibinfo
  {author} {\bibfnamefont {S.~A.}\ \bibnamefont {Mikhailov}},\ }\bibfield
  {title} {\enquote {\bibinfo {title} {{Coherent Nonlinear Optical Response of
  Graphene}},}\ }\href {\doibase 10.1103/PhysRevLett.105.097401} {\bibfield
  {journal} {\bibinfo  {journal} {Physical Review Letters}\ }\textbf {\bibinfo
  {volume} {105}},\ \bibinfo {pages} {097401} (\bibinfo {year}
  {2010})}\BibitemShut {NoStop}%
\bibitem [{\citenamefont {Gullans}\ \emph {et~al.}(2013)\citenamefont
  {Gullans}, \citenamefont {Chang}, \citenamefont {Koppens}, \citenamefont
  {de~Abajo},\ and\ \citenamefont {Lukin}}]{Gullans2013}%
  \BibitemOpen
  \bibfield  {author} {\bibinfo {author} {\bibfnamefont {M.}~\bibnamefont
  {Gullans}}, \bibinfo {author} {\bibfnamefont {D.}~\bibnamefont {Chang}},
  \bibinfo {author} {\bibfnamefont {F.}~\bibnamefont {Koppens}}, \bibinfo
  {author} {\bibfnamefont {F.}~\bibnamefont {de~Abajo}}, \ and\ \bibinfo
  {author} {\bibfnamefont {M.}~\bibnamefont {Lukin}},\ }\bibfield  {title}
  {\enquote {\bibinfo {title} {{Single-Photon Nonlinear Optics with Graphene
  Plasmons}},}\ }\href {\doibase 10.1103/PhysRevLett.111.247401} {\bibfield
  {journal} {\bibinfo  {journal} {Physical Review Letters}\ }\textbf {\bibinfo
  {volume} {111}},\ \bibinfo {pages} {247401} (\bibinfo {year}
  {2013})}\BibitemShut {NoStop}%
\bibitem [{\citenamefont {Wang}\ \emph
  {et~al.}(2013{\natexlab{a}})\citenamefont {Wang}, \citenamefont {Meric},
  \citenamefont {Huang}, \citenamefont {Gao}, \citenamefont {Gao},
  \citenamefont {Tran}, \citenamefont {Taniguchi}, \citenamefont {Watanabe},
  \citenamefont {Campos}, \citenamefont {Muller}, \citenamefont {Guo},
  \citenamefont {Kim}, \citenamefont {Hone}, \citenamefont {Shepard},\ and\
  \citenamefont {Dean}}]{Wang2013b}%
  \BibitemOpen
  \bibfield  {author} {\bibinfo {author} {\bibfnamefont {L.}~\bibnamefont
  {Wang}}, \bibinfo {author} {\bibfnamefont {I.}~\bibnamefont {Meric}},
  \bibinfo {author} {\bibfnamefont {P.~Y.}\ \bibnamefont {Huang}}, \bibinfo
  {author} {\bibfnamefont {Q.}~\bibnamefont {Gao}}, \bibinfo {author}
  {\bibfnamefont {Y.}~\bibnamefont {Gao}}, \bibinfo {author} {\bibfnamefont
  {H.}~\bibnamefont {Tran}}, \bibinfo {author} {\bibfnamefont {T.}~\bibnamefont
  {Taniguchi}}, \bibinfo {author} {\bibfnamefont {K.}~\bibnamefont {Watanabe}},
  \bibinfo {author} {\bibfnamefont {L.~M.}\ \bibnamefont {Campos}}, \bibinfo
  {author} {\bibfnamefont {D.~a.}\ \bibnamefont {Muller}}, \bibinfo {author}
  {\bibfnamefont {J.}~\bibnamefont {Guo}}, \bibinfo {author} {\bibfnamefont
  {P.}~\bibnamefont {Kim}}, \bibinfo {author} {\bibfnamefont {J.}~\bibnamefont
  {Hone}}, \bibinfo {author} {\bibfnamefont {K.~L.}\ \bibnamefont {Shepard}}, \
  and\ \bibinfo {author} {\bibfnamefont {C.~R.}\ \bibnamefont {Dean}},\
  }\bibfield  {title} {\enquote {\bibinfo {title} {{One-dimensional electrical
  contact to a two-dimensional material.}}}\ }\href {\doibase
  10.1126/science.1244358} {\bibfield  {journal} {\bibinfo  {journal} {Science
  (New York, N.Y.)}\ }\textbf {\bibinfo {volume} {342}},\ \bibinfo {pages}
  {614--7} (\bibinfo {year} {2013}{\natexlab{a}})}\BibitemShut {NoStop}%
\bibitem [{\citenamefont {Geim}\ and\ \citenamefont
  {Grigorieva}(2013)}]{Geim2013}%
  \BibitemOpen
  \bibfield  {author} {\bibinfo {author} {\bibfnamefont {A.~K.}\ \bibnamefont
  {Geim}}\ and\ \bibinfo {author} {\bibfnamefont {I.~V.}\ \bibnamefont
  {Grigorieva}},\ }\bibfield  {title} {{\selectlanguage {english}\enquote {\bibinfo
  {title} {{Van der Waals heterostructures.}}}\ }}\href {\doibase
  10.1038/nature12385} {\bibfield  {journal} {\bibinfo  {journal} {Nature}\
  }\textbf {\bibinfo {volume} {499}},\ \bibinfo {pages} {419--25} (\bibinfo
  {year} {2013})}\BibitemShut {NoStop}%
\bibitem [{\citenamefont {Song}\ \emph {et~al.}(2011)\citenamefont {Song},
  \citenamefont {Rudner}, \citenamefont {Marcus},\ and\ \citenamefont
  {Levitov}}]{Song2011b}%
  \BibitemOpen
  \bibfield  {author} {\bibinfo {author} {\bibfnamefont {J.~C.~W.}\
  \bibnamefont {Song}}, \bibinfo {author} {\bibfnamefont {M.~S.}\ \bibnamefont
  {Rudner}}, \bibinfo {author} {\bibfnamefont {C.~M.}\ \bibnamefont {Marcus}},
  \ and\ \bibinfo {author} {\bibfnamefont {L.~S.}\ \bibnamefont {Levitov}},\
  }\bibfield  {title} {\enquote {\bibinfo {title} {{Hot carrier transport and
  photocurrent response in graphene.}}}\ }\href {\doibase 10.1021/nl202318u}
  {\bibfield  {journal} {\bibinfo  {journal} {Nano letters}\ }\textbf {\bibinfo
  {volume} {11}},\ \bibinfo {pages} {4688--4692} (\bibinfo {year}
  {2011})}\BibitemShut {NoStop}%
\bibitem [{\citenamefont {Song}, \citenamefont {Reizer},\ and\ \citenamefont
  {Levitov}(2012)}]{Song2012}%
  \BibitemOpen
  \bibfield  {author} {\bibinfo {author} {\bibfnamefont {J.~C.~W.}\
  \bibnamefont {Song}}, \bibinfo {author} {\bibfnamefont {M.~Y.}\ \bibnamefont
  {Reizer}}, \ and\ \bibinfo {author} {\bibfnamefont {L.~S.}\ \bibnamefont
  {Levitov}},\ }\bibfield  {title} {\enquote {\bibinfo {title}
  {{Disorder-Assisted Electron-Phonon Scattering and Cooling Pathways in
  Graphene}},}\ }\href {\doibase 10.1103/PhysRevLett.109.106602} {\bibfield
  {journal} {\bibinfo  {journal} {Physical Review Letters}\ }\textbf {\bibinfo
  {volume} {109}},\ \bibinfo {pages} {106602} (\bibinfo {year}
  {2012})}\BibitemShut {NoStop}%
\bibitem [{\citenamefont {Graham}\ \emph {et~al.}(2012)\citenamefont {Graham},
  \citenamefont {Shi}, \citenamefont {Ralph}, \citenamefont {Park},\ and\
  \citenamefont {McEuen}}]{Graham2012}%
  \BibitemOpen
  \bibfield  {author} {\bibinfo {author} {\bibfnamefont {M.~W.}\ \bibnamefont
  {Graham}}, \bibinfo {author} {\bibfnamefont {S.-F.}\ \bibnamefont {Shi}},
  \bibinfo {author} {\bibfnamefont {D.~C.}\ \bibnamefont {Ralph}}, \bibinfo
  {author} {\bibfnamefont {J.}~\bibnamefont {Park}}, \ and\ \bibinfo {author}
  {\bibfnamefont {P.~L.}\ \bibnamefont {McEuen}},\ }\bibfield  {title}
  {\enquote {\bibinfo {title} {{Photocurrent measurements of supercollision
  cooling in graphene}},}\ }\href {\doibase 10.1038/nphys2493} {\bibfield
  {journal} {\bibinfo  {journal} {Nature Physics}\ }\textbf {\bibinfo {volume}
  {9}},\ \bibinfo {pages} {103--108} (\bibinfo {year} {2012})}\BibitemShut
  {NoStop}%
\bibitem [{\citenamefont {Foster}\ \emph {et~al.}(2008)\citenamefont {Foster},
  \citenamefont {Salem}, \citenamefont {Geraghty}, \citenamefont
  {Turner-Foster}, \citenamefont {Lipson},\ and\ \citenamefont
  {Gaeta}}]{Foster2008}%
  \BibitemOpen
  \bibfield  {author} {\bibinfo {author} {\bibfnamefont {M.~a.}\ \bibnamefont
  {Foster}}, \bibinfo {author} {\bibfnamefont {R.}~\bibnamefont {Salem}},
  \bibinfo {author} {\bibfnamefont {D.~F.}\ \bibnamefont {Geraghty}}, \bibinfo
  {author} {\bibfnamefont {A.~C.}\ \bibnamefont {Turner-Foster}}, \bibinfo
  {author} {\bibfnamefont {M.}~\bibnamefont {Lipson}}, \ and\ \bibinfo {author}
  {\bibfnamefont {A.~L.}\ \bibnamefont {Gaeta}},\ }\bibfield  {title} {\enquote
  {\bibinfo {title} {{Silicon-chip-based ultrafast optical oscilloscope.}}}\
  }\href {\doibase 10.1038/nature07430} {\bibfield  {journal} {\bibinfo
  {journal} {Nature}\ }\textbf {\bibinfo {volume} {456}},\ \bibinfo {pages}
  {81--4} (\bibinfo {year} {2008})}\BibitemShut {NoStop}%
\bibitem [{\citenamefont {Tien}\ \emph {et~al.}(2009)\citenamefont {Tien},
  \citenamefont {Sang}, \citenamefont {Qing}, \citenamefont {Song},\ and\
  \citenamefont {Boyraz}}]{Tien2009}%
  \BibitemOpen
  \bibfield  {author} {\bibinfo {author} {\bibfnamefont {E.-K.}\ \bibnamefont
  {Tien}}, \bibinfo {author} {\bibfnamefont {X.-Z.}\ \bibnamefont {Sang}},
  \bibinfo {author} {\bibfnamefont {F.}~\bibnamefont {Qing}}, \bibinfo {author}
  {\bibfnamefont {Q.}~\bibnamefont {Song}}, \ and\ \bibinfo {author}
  {\bibfnamefont {O.}~\bibnamefont {Boyraz}},\ }\bibfield  {title} {\enquote
  {\bibinfo {title} {{Ultrafast pulse characterization using cross phase
  modulation in silicon}},}\ }\href {\doibase 10.1063/1.3193538} {\bibfield
  {journal} {\bibinfo  {journal} {Applied Physics Letters}\ }\textbf {\bibinfo
  {volume} {95}},\ \bibinfo {pages} {051101} (\bibinfo {year}
  {2009})}\BibitemShut {NoStop}%
\bibitem [{\citenamefont {Walmsley}\ and\ \citenamefont
  {Dorrer}(2009)}]{Walmsley2009}%
  \BibitemOpen
  \bibfield  {author} {\bibinfo {author} {\bibfnamefont {I.~A.}\ \bibnamefont
  {Walmsley}}\ and\ \bibinfo {author} {\bibfnamefont {C.}~\bibnamefont
  {Dorrer}},\ }\bibfield  {title} {\enquote {\bibinfo {title}
  {{Characterization of ultrashort electromagnetic pulses}},}\ }\href {\doibase
  10.1364/AOP.1.000308} {\bibfield  {journal} {\bibinfo  {journal} {Advances in
  Optics and Photonics}\ }\textbf {\bibinfo {volume} {1}},\ \bibinfo {pages}
  {308} (\bibinfo {year} {2009})}\BibitemShut {NoStop}%
\bibitem [{\citenamefont {Lu}\ \emph {et~al.}(2004)\citenamefont {Lu},
  \citenamefont {Fu}, \citenamefont {Huang},\ and\ \citenamefont
  {Liu}}]{Lu2004}%
  \BibitemOpen
  \bibfield  {author} {\bibinfo {author} {\bibfnamefont {C.}~\bibnamefont
  {Lu}}, \bibinfo {author} {\bibfnamefont {Q.}~\bibnamefont {Fu}}, \bibinfo
  {author} {\bibfnamefont {S.}~\bibnamefont {Huang}}, \ and\ \bibinfo {author}
  {\bibfnamefont {J.}~\bibnamefont {Liu}},\ }\bibfield  {title} {\enquote
  {\bibinfo {title} {{Polymer Electrolyte-Gated Carbon Nanotube Field-Effect
  Transistor}},}\ }\href {\doibase 10.1021/nl049937e} {\bibfield  {journal}
  {\bibinfo  {journal} {Nano Letters}\ }\textbf {\bibinfo {volume} {4}},\
  \bibinfo {pages} {623--627} (\bibinfo {year} {2004})}\BibitemShut {NoStop}%
\bibitem [{\citenamefont {Pospischil}\ \emph {et~al.}(2013)\citenamefont
  {Pospischil}, \citenamefont {Humer}, \citenamefont {Furchi}, \citenamefont
  {Bachmann}, \citenamefont {Guider}, \citenamefont {Fromherz},\ and\
  \citenamefont {Mueller}}]{Pospischil2013}%
  \BibitemOpen
  \bibfield  {author} {\bibinfo {author} {\bibfnamefont {A.}~\bibnamefont
  {Pospischil}}, \bibinfo {author} {\bibfnamefont {M.}~\bibnamefont {Humer}},
  \bibinfo {author} {\bibfnamefont {M.~M.}\ \bibnamefont {Furchi}}, \bibinfo
  {author} {\bibfnamefont {D.}~\bibnamefont {Bachmann}}, \bibinfo {author}
  {\bibfnamefont {R.}~\bibnamefont {Guider}}, \bibinfo {author} {\bibfnamefont
  {T.}~\bibnamefont {Fromherz}}, \ and\ \bibinfo {author} {\bibfnamefont
  {T.}~\bibnamefont {Mueller}},\ }\bibfield  {title} {{\selectlanguage
  {english}\enquote {\bibinfo {title} {{CMOS-compatible graphene photodetector
  covering all optical communication bands}},}\ }}\href {\doibase
  10.1038/nphoton.2013.240} {\bibfield  {journal} {\bibinfo  {journal} {Nature
  Photonics}\ }\textbf {\bibinfo {volume} {7}},\ \bibinfo {pages} {892--896}
  (\bibinfo {year} {2013})}\BibitemShut {NoStop}%
\bibitem [{\citenamefont {Gan}\ \emph {et~al.}(2013{\natexlab{b}})\citenamefont
  {Gan}, \citenamefont {Shiue}, \citenamefont {Gao}, \citenamefont {Meric},
  \citenamefont {Heinz}, \citenamefont {Shepard}, \citenamefont {Hone},
  \citenamefont {Assefa},\ and\ \citenamefont {Englund}}]{Gan2013}%
  \BibitemOpen
  \bibfield  {author} {\bibinfo {author} {\bibfnamefont {X.}~\bibnamefont
  {Gan}}, \bibinfo {author} {\bibfnamefont {R.-J.}\ \bibnamefont {Shiue}},
  \bibinfo {author} {\bibfnamefont {Y.}~\bibnamefont {Gao}}, \bibinfo {author}
  {\bibfnamefont {I.}~\bibnamefont {Meric}}, \bibinfo {author} {\bibfnamefont
  {T.~F.}\ \bibnamefont {Heinz}}, \bibinfo {author} {\bibfnamefont
  {K.}~\bibnamefont {Shepard}}, \bibinfo {author} {\bibfnamefont
  {J.}~\bibnamefont {Hone}}, \bibinfo {author} {\bibfnamefont {S.}~\bibnamefont
  {Assefa}}, \ and\ \bibinfo {author} {\bibfnamefont {D.}~\bibnamefont
  {Englund}},\ }\bibfield  {title} {\enquote {\bibinfo {title}
  {{Chip-integrated ultrafast graphene photodetector with high
  responsivity}},}\ }\href {\doibase 10.1038/nphoton.2013.253} {\bibfield
  {journal} {\bibinfo  {journal} {Nature Photonics}\ }\textbf {\bibinfo
  {volume} {7}},\ \bibinfo {pages} {883--887} (\bibinfo {year}
  {2013}{\natexlab{b}})}\BibitemShut {NoStop}%
\bibitem [{\citenamefont {Wang}\ \emph
  {et~al.}(2013{\natexlab{b}})\citenamefont {Wang}, \citenamefont {Cheng},
  \citenamefont {Xu}, \citenamefont {Tsang},\ and\ \citenamefont
  {Xu}}]{Wang2013}%
  \BibitemOpen
  \bibfield  {author} {\bibinfo {author} {\bibfnamefont {X.}~\bibnamefont
  {Wang}}, \bibinfo {author} {\bibfnamefont {Z.}~\bibnamefont {Cheng}},
  \bibinfo {author} {\bibfnamefont {K.}~\bibnamefont {Xu}}, \bibinfo {author}
  {\bibfnamefont {H.~K.}\ \bibnamefont {Tsang}}, \ and\ \bibinfo {author}
  {\bibfnamefont {J.-B.}\ \bibnamefont {Xu}},\ }\bibfield  {title} {\enquote
  {\bibinfo {title} {{High-responsivity graphene/silicon-heterostructure
  waveguide photodetectors}},}\ }\href {\doibase 10.1038/nphoton.2013.241}
  {\bibfield  {journal} {\bibinfo  {journal} {Nature Photonics}\ }\textbf
  {\bibinfo {volume} {7}},\ \bibinfo {pages} {888--891} (\bibinfo {year}
  {2013}{\natexlab{b}})}\BibitemShut {NoStop}%
\bibitem [{\citenamefont {Gabor}\ \emph {et~al.}(2011)\citenamefont {Gabor},
  \citenamefont {Song}, \citenamefont {Ma}, \citenamefont {Nair}, \citenamefont
  {Taychatanapat}, \citenamefont {Watanabe}, \citenamefont {Taniguchi},
  \citenamefont {Levitov},\ and\ \citenamefont {Jarillo-Herrero}}]{Gabor2011c}%
  \BibitemOpen
  \bibfield  {author} {\bibinfo {author} {\bibfnamefont {N.~M.}\ \bibnamefont
  {Gabor}}, \bibinfo {author} {\bibfnamefont {J.~C.~W.}\ \bibnamefont {Song}},
  \bibinfo {author} {\bibfnamefont {Q.}~\bibnamefont {Ma}}, \bibinfo {author}
  {\bibfnamefont {N.~L.}\ \bibnamefont {Nair}}, \bibinfo {author}
  {\bibfnamefont {T.}~\bibnamefont {Taychatanapat}}, \bibinfo {author}
  {\bibfnamefont {K.}~\bibnamefont {Watanabe}}, \bibinfo {author}
  {\bibfnamefont {T.}~\bibnamefont {Taniguchi}}, \bibinfo {author}
  {\bibfnamefont {L.~S.}\ \bibnamefont {Levitov}}, \ and\ \bibinfo {author}
  {\bibfnamefont {P.}~\bibnamefont {Jarillo-Herrero}},\ }\bibfield  {title}
  {\enquote {\bibinfo {title} {{Hot carrier-assisted intrinsic photoresponse in
  graphene.}}}\ }\href {\doibase 10.1126/science.1211384} {\bibfield  {journal}
  {\bibinfo  {journal} {Science (New York, N.Y.)}\ }\textbf {\bibinfo {volume}
  {334}},\ \bibinfo {pages} {648--52} (\bibinfo {year} {2011})}\BibitemShut
  {NoStop}%
\bibitem [{\citenamefont {Ashcroft}\ and\ \citenamefont
  {Mermin}(1976)}]{Ashcroft1976}%
  \BibitemOpen
  \bibfield  {author} {\bibinfo {author} {\bibfnamefont {N.~W.}\ \bibnamefont
  {Ashcroft}}\ and\ \bibinfo {author} {\bibfnamefont {N.~D.}\ \bibnamefont
  {Mermin}},\ }\href
  {http://books.google.com/books/about/Solid\_State\_Physics.html?id=oXIfAQAAMAAJ\&pgis=1}
  {\emph {\bibinfo {title} {{Solid State Physics}}}}\ (\bibinfo  {publisher}
  {Holt, Rinehart and Winston},\ \bibinfo {year} {1976})\ p.\ \bibinfo {pages}
  {826}\BibitemShut {NoStop}%
\bibitem [{\citenamefont {Liu}\ \emph {et~al.}(2012)\citenamefont {Liu},
  \citenamefont {Dissanayake}, \citenamefont {Lee}, \citenamefont {Lee},\ and\
  \citenamefont {Zhong}}]{Liu2012}%
  \BibitemOpen
  \bibfield  {author} {\bibinfo {author} {\bibfnamefont {C.-H.}\ \bibnamefont
  {Liu}}, \bibinfo {author} {\bibfnamefont {N.~M.}\ \bibnamefont
  {Dissanayake}}, \bibinfo {author} {\bibfnamefont {S.}~\bibnamefont {Lee}},
  \bibinfo {author} {\bibfnamefont {K.}~\bibnamefont {Lee}}, \ and\ \bibinfo
  {author} {\bibfnamefont {Z.}~\bibnamefont {Zhong}},\ }\bibfield  {title}
  {\enquote {\bibinfo {title} {{Evidence for extraction of photoexcited hot
  carriers from graphene.}}}\ }\href {\doibase 10.1021/nn302227r} {\bibfield
  {journal} {\bibinfo  {journal} {ACS nano}\ }\textbf {\bibinfo {volume} {6}},\
  \bibinfo {pages} {7172--6} (\bibinfo {year} {2012})}\BibitemShut {NoStop}%
\bibitem [{\citenamefont {Lemme}\ \emph {et~al.}(2011)\citenamefont {Lemme},
  \citenamefont {Koppens}, \citenamefont {Falk}, \citenamefont {Rudner},
  \citenamefont {Park}, \citenamefont {Levitov},\ and\ \citenamefont
  {Marcus}}]{Lemme2011}%
  \BibitemOpen
  \bibfield  {author} {\bibinfo {author} {\bibfnamefont {M.~C.}\ \bibnamefont
  {Lemme}}, \bibinfo {author} {\bibfnamefont {F.~H.~L.}\ \bibnamefont
  {Koppens}}, \bibinfo {author} {\bibfnamefont {A.~L.}\ \bibnamefont {Falk}},
  \bibinfo {author} {\bibfnamefont {M.~S.}\ \bibnamefont {Rudner}}, \bibinfo
  {author} {\bibfnamefont {H.}~\bibnamefont {Park}}, \bibinfo {author}
  {\bibfnamefont {L.~S.}\ \bibnamefont {Levitov}}, \ and\ \bibinfo {author}
  {\bibfnamefont {C.~M.}\ \bibnamefont {Marcus}},\ }\bibfield  {title}
  {\enquote {\bibinfo {title} {{Gate-activated photoresponse in a graphene p-n
  junction.}}}\ }\href {\doibase 10.1021/nl2019068} {\bibfield  {journal}
  {\bibinfo  {journal} {Nano letters}\ }\textbf {\bibinfo {volume} {11}},\
  \bibinfo {pages} {4134--4137} (\bibinfo {year} {2011})}\BibitemShut {NoStop}%
\bibitem [{Note1()}]{Note1}%
  \BibitemOpen
  \bibinfo {note} {Reef femtosecond autocorrelator Reef-RTD and Femtochrome
  Research Inc. FR-103WS}\BibitemShut {NoStop}%
\bibitem [{\citenamefont {Hayat}\ \emph {et~al.}(2011)\citenamefont {Hayat},
  \citenamefont {Nevet}, \citenamefont {Ginzburg},\ and\ \citenamefont
  {Orenstein}}]{Hayat2011}%
  \BibitemOpen
  \bibfield  {author} {\bibinfo {author} {\bibfnamefont {A.}~\bibnamefont
  {Hayat}}, \bibinfo {author} {\bibfnamefont {A.}~\bibnamefont {Nevet}},
  \bibinfo {author} {\bibfnamefont {P.}~\bibnamefont {Ginzburg}}, \ and\
  \bibinfo {author} {\bibfnamefont {M.}~\bibnamefont {Orenstein}},\ }\bibfield
  {title} {\enquote {\bibinfo {title} {{Applications of two-photon processes in
  semiconductor photonic devices: invited review}},}\ }\href {\doibase
  10.1088/0268-1242/26/8/083001} {\bibfield  {journal} {\bibinfo  {journal}
  {Semiconductor Science and Technology}\ }\textbf {\bibinfo {volume} {26}},\
  \bibinfo {pages} {083001} (\bibinfo {year} {2011})}\BibitemShut {NoStop}%
\bibitem [{\citenamefont {Duchesne}\ \emph {et~al.}(2010)\citenamefont
  {Duchesne}, \citenamefont {Razzari}, \citenamefont {Halloran},\ and\
  \citenamefont {Gigu}}]{Duchesne2010}%
  \BibitemOpen
  \bibfield  {author} {\bibinfo {author} {\bibfnamefont {D.}~\bibnamefont
  {Duchesne}}, \bibinfo {author} {\bibfnamefont {L.}~\bibnamefont {Razzari}},
  \bibinfo {author} {\bibfnamefont {L.}~\bibnamefont {Halloran}}, \ and\
  \bibinfo {author} {\bibfnamefont {M.}~\bibnamefont {Gigu}},\ }\bibfield
  {title} {\enquote {\bibinfo {title} {{Two-photon Autocorrelation in a MQW
  GaAs Laser at 1 . 55 µ m}},}\ }\href@noop {} {\ \textbf {\bibinfo {volume}
  {0}},\ \bibinfo {pages} {267--272} (\bibinfo {year} {2010})}\BibitemShut
  {NoStop}%
\bibitem [{\citenamefont {Liang}\ \emph {et~al.}(2002)\citenamefont {Liang},
  \citenamefont {Tsang}, \citenamefont {Day}, \citenamefont {Drake},
  \citenamefont {Knights},\ and\ \citenamefont {Asghari}}]{Liang2002}%
  \BibitemOpen
  \bibfield  {author} {\bibinfo {author} {\bibfnamefont {T.~K.}\ \bibnamefont
  {Liang}}, \bibinfo {author} {\bibfnamefont {H.~K.}\ \bibnamefont {Tsang}},
  \bibinfo {author} {\bibfnamefont {I.~E.}\ \bibnamefont {Day}}, \bibinfo
  {author} {\bibfnamefont {J.}~\bibnamefont {Drake}}, \bibinfo {author}
  {\bibfnamefont {A.~P.}\ \bibnamefont {Knights}}, \ and\ \bibinfo {author}
  {\bibfnamefont {M.}~\bibnamefont {Asghari}},\ }\bibfield  {title} {\enquote
  {\bibinfo {title} {{Silicon waveguide two-photon absorption detector at 1.5
  $\mu$m wavelength for autocorrelation measurements}},}\ }\href {\doibase
  10.1063/1.1500430} {\bibfield  {journal} {\bibinfo  {journal} {Applied
  Physics Letters}\ }\textbf {\bibinfo {volume} {81}},\ \bibinfo {pages} {1323}
  (\bibinfo {year} {2002})}\BibitemShut {NoStop}%
\end{thebibliography}
\end{document}